\documentstyle[12pt,epsfig,amsfonts]{article}
 
\advance \topmargin by -\headheight
\advance \topmargin by -10mm
 
\evensidemargin \oddsidemargin
\newcommand{\R}{{\bf R}}

\newcommand{\CP}{{\bf CP}}
\newcommand{\C}{{\bf C}}

\newcommand{\sigmabf}{\mbox{\boldmath $\sigma$}}

\begin{document}
\title{
  \vskip 15pt
  {\bf \Large \bf Folding in the Skyrme Model}
  \vskip 10pt}
\author{ 
  Conor J. Houghton\thanks{houghton@maths.tcd.ie}
\\[5pt] 
{\normalsize {\sl School of Mathematics}}\\{\normalsize {\sl Trinity
College, Dublin 2}}\\{\normalsize {\sl Ireland}}\\[10pt] 
and
Steffen Krusch\thanks{s.krusch@damtp.cam.ac.uk } \\[5pt]
{\normalsize {\sl Department of Applied Mathematics
      and Theoretical Physics}}\\
{\normalsize {\sl Centre for Mathematical Sciences, Wilberforce Road,
Cambridge CB3 0WA}}\\ 
{\normalsize {\sl England}}
}

\date{April 2001}
 
\maketitle

\begin{abstract}

There are only three stable singularities of a differentiable map
between three-dimensional manifolds, namely folds, cusps and
swallowtails. A Skyrme configuration is a map from space to
SU$_2$, and its singularities correspond to the points where the baryon
density vanishes. In this paper we consider the  singularity
structure of Skyrme configurations.

The Skyrme model can only be solved numerically.  However, there are
good analytic ans\"atze.  The simplest of these, the rational map
ansatz, has a non-generic singularity structure. This leads us to
introduce a non-holomorphic ansatz as a
generalization. For baryon number two, three and four, the approximate
solutions derived from this ansatz are closer in energy to the true
solutions than any other ansatz solution. We find that there is a tiny
amount of negative baryon density for baryon number three, but none
for two or four. We comment briefly on the relationship to
Bogomolny-Prasad-Sommerfield monopoles. 

\end{abstract}

\newpage
\section{Introduction}

The Skyrme model is a nonlinear SU$_2$ field theory
\cite{Skyrme:1961vq}.  In addition to the fundamental excitations, the
spectrum also includes topologically-charged soliton solutions. The model
was proposed by Skyrme as a theory of nuclear physics in which the
fundamental excitations are pions and the solitons are nucleons. The
Skyrme energy function is
\begin{equation}
E=\int\left\{-\frac{1}{2}\mbox{Tr}\,(R_iR_i)- 
\frac{1}{16}\mbox{Tr}([R_i,R_j][R_i,R_j])\right\}\mbox{d}^3x,
\end{equation}
where $R_i$ is the $su_2$-valued current $R_i=(\partial_iU)U^{-1}$. 
The SU$_2$-valued field $U$ is required to attain its vacuum value,
the identity, at spatial infinity and, so, it is a map between
topological three-spheres. This is the origin of the topological charge,
$B$. 

The 1-Skyrmion is spherical and is given by the hedgehog ansatz
\begin{equation}
U_1({\bf x})=\exp(if(r) \ \widehat{\bf n}\cdot\sigmabf),\label{HH}
\end{equation}
where $\widehat{\bf n} = \widehat{\bf x}$ 
is the outward pointing unit normal and 
$\sigmabf=(\sigma_1,\sigma_2,\sigma_3)$ are the Pauli matrices. $f(r)$
is a shape function and is usually determined numerically. It is
very well approximated by the kink profile \cite{Sutcliffe:1992bg}:
$f(r)=4\arctan({\exp(-r)})$.  The 1-Skyrmion has six zero modes: three
translational modes and three isospin modes corresponding to global
SU$_2$ transformations.

Two well-separated Skyrmions attract or repel, depending upon their
mutual isospin orientation. Two attracting 1-Skyrmions will move towards
each other and form a bound state whose energy is $0.95$ times the
energy of two 1-Skyrmions.  This 2-Skyrmion is torus shaped
\cite{Kopeliovich:1987bt, Verbaarschot:1987au}
and is axially symmetric, in the sense that axial rotations in space
are equivalent to isospin rotations, which are conjugations of $U$ by a
constant SU$_2$ matrix.

In the Skyrme model, the classical $B$-nucleon nucleus is a
$B$-Skyrmion; that is, a minimum energy Skyrme field with topological
charge $B$.  For $B$ from three to 22, the Skyrmion has been
calculated numerically by evolving an attractive configuration,
\cite{Braaten:1990rg,Battye:1997qq, Battye:2000se}. All the known
Skyrmions have the bulk of their energy density on a fullerene-like
shell.  A geometric interpretation of this shell-like structure was
given in \cite{Krusch:2000gb}.  Furthermore, Skyrmions often have a
very symmetrical shape.

\subsection{The Rational Map Ansatz}

The rational map ansatz introduced in \cite{Houghton:1998kg} is a
simple ansatz for Skyrmions. It is similar to the 1-Skyrmion 
(\ref{HH}).  The 1-Skyrmion is a hedgehog map in which the outward
pointing unit normal, $\widehat{\bf n}$, 
maps a two-sphere identically to a two-sphere. In
the ansatz, the hedgehog map is replaced by a more general holomorphic
map, $\widehat{\bf n}_R$, from Riemann sphere to Riemann sphere.  
The rational map ansatz is given by
\begin{equation}
U(r,z)=\exp(if(r) \ \widehat{\bf n}_R\cdot\sigmabf),
\label{ansatz}
\end{equation}
where
\begin{equation} 
\widehat{\bf n}_R=\frac{1}{1+\vert
R\vert^2}(2\mbox{Re}(R), 2\mbox{Im}(R),1-\vert R\vert^2)
\end{equation}
and $R(z)$ is a holomorphic map in $z$.  Here, $z$ is related to the
standard angular coordinates $\phi$ and $\theta$ by the stereographic
projection $z = \tan (\theta/2) \exp(i \phi)$.  $R$ is also
a stereographic coordinate on the Riemann sphere.  In the ansatz, this
Riemann sphere is a latitudinal two-sphere in SU$_2\cong S^3$. 

The ansatz maps spheres around the origin in space to
latitudinal two-spheres in SU$_2$. The shape function $f$ is a function
of $r$ only, so each map between two-spheres is identical. The
boundary conditions on $f$ are $f(0)=\pi$ and
$f(\infty)=0$. These conditions are determined by requiring that $U$
is well defined at the origin and attains the vacuum value at
infinity. In principle, we could have $f(0)=N_f\pi$ for any non-zero
integer $N_f$, but solutions with $N_f>1$ have rather high energy and
so we only consider $N_f=1$. 

Thus, the ansatz depends on a holomorphic map between two-spheres. Any
holomorphic map of finite topological charge can be written as a
rational map
\begin{equation}
\label{R=p/q}
R(z) = \frac{p(z)}{q(z)},
\end{equation}
where $p$ and $q$ are polynomials in $z$. The
topological charge, $N_R$, of the map is equal to the algebraic degree
and  this, in turn, is given by the maximal degree of the two
polynomials.

The easiest way to calculate the energy of an ansatz field is to use
the geometric formulation of the Skyrme model \cite{Manton:1987xt}.
A Skyrme field is a
map between three-manifolds with metrics and so there is a strain
tensor. This is given by
\begin{equation}
\label{straintensor}
D_{ij}= - \frac{1}{2} {\rm Tr} \left( R_i R_j \right).
\end{equation}
The static energy, $E$, and the baryon number, $B$, of a Skyrme field can
be written in terms of the eigenvalues, $\lambda_1^2$, $\lambda_2^2$
and $\lambda_3^2$, of this tensor: 
\begin{eqnarray}
\label{Egeo}
E &=& \int \left(\lambda_1^2+\lambda_2^2+\lambda_3^2 +
\lambda_1^2\lambda_2^2+\lambda_2^2\lambda_3^2+\lambda_1^2\lambda_3^2\right),
\\
\label{Bgeo}
B &=& \frac{1}{2 \pi^2} \int \lambda_1 \lambda_2 \lambda_3.
\end{eqnarray}
The strain tensor of the ansatz field (\ref{ansatz}) can be calculated
to give 
\begin{eqnarray}
\label{Brational}
B &=& \frac{2}{\pi} \int \left(
   -f^{\prime} \sin^2 f \right) {\rm d} r~
   \frac{1}{4 \pi} \int \left(
    \frac{1+|z|^2}{1+|R|^2}\left| \frac{{\rm d} R}{{\rm d} z} \right|
    \right)^2 \frac{2i~{\rm d}z {\rm d}{\bar z}}{(1+ |z|^2)^2},  \\
\nonumber \\
\label{B=NfNR}
  &=& N_R,
\end{eqnarray}
where $N_R$ is the degree of the rational map.

Similarly, the energy $E$ is given by
\begin{equation}
\label{Erational}
E = 4 \pi \int
\left(
f^{\prime 2} r^2
 + 2 N_R (f^{\prime 2} + 1) \sin^2 f +
{\cal I}~ \frac{\sin^4 f}{r^2}
\right)
{\rm d} r,
\end{equation}  
where 
\begin{equation}
\label{calI}
{\cal I}  = \frac{1}{4 \pi} \int \left(
\frac{1 + |z|^2}{1 + |R|^2} \left|\frac{{\rm d} R}{{\rm d} z} \right|
\right)^4
\frac{2i~{\rm d} z {\rm d}{\bar z}}{(1 + |z|^2)^2}.
\end{equation}
Thus, the minimum energy ansatz field is found by choosing polynomials
$p$ and $q$ which minimize ${\cal I}$ and then calculating the shape
function $f$ numerically. These minimum energy ansatz fields have been
calculated for all the known Skyrmions and are found to have energies
that exceed the true, numerically determined,
minima \cite{Battye:2000se, Houghton:1998kg} by less than three
percent.

\subsection{The Rational Map Ansatz and Negative Baryon Density}

The accuracy of the ansatz is established by observing how close in
energy the minimum energy ansatz configurations are to the true
minima. In other words, it is not known whether the ansatz fields
resemble the true fields in regions where the energy density is
low. 
In fact, for the approximate fields calculated within the ansatz, the
region of zero baryon density has a rather special structure.
There are $2B-2$ radial half-lines which meet at the
origin and extend out to infinity.
The zeros of the baryon density correspond to the folds
in the Skyrme fields, considered as maps between three-spheres.
Line-like zeros are not generic.  It is possible that these
non-generic zeros are a natural consequence of minimizing the Skyrme
energy \cite{Baskerville:1999kk, Kugler:1997pz}. 
However, it may be that this is a weakness of the rational map ansatz
and Skyrmions have a more generic folding structure.

If it is a weakness, it is not a very
serious one. Most current interest is in finding minimum energy Skyrme
configurations. However, the issue of determining the structure of
Skyrme fields in the regions where there is little energy may be of
some practical importance, for example, in calculations in which the
Skyrmion fields are used as backgrounds for fermion excitations
modeling heavy flavours, see \cite{Schat:1999dw}.

In this paper, we consider the consequences of generalizing the
rational map ansatz to include a larger class of maps. These maps 
permit a more natural, though not wholly generic, folding
structure. This generalized ansatz is not as convenient as the
original one. However, it does result in ansatz fields which are even
closer to the true minima.

Our interest in this problem is partly motivated by BPS monopoles.
There are many interesting similarities between Skyrmions and BPS
monopoles. For example, there are two rational map descriptions of
monopoles \cite{Hurtubise:1985vq, Jarvis:1996}, 
and it is widely believed that the space of
attracting Skyrmions is related to the space of monopoles. It was
discovered in \cite{Houghton:1996bs} that the tetrahedral 3-monopole
has a negative multiplicity Higgs zero. Subsequent examination
revealed that the octahedral 5-monopole also has extra zeros but the
cubic 4-monopole does not \cite{Sutcliffe:1996qz}. 

There is evidence that this pattern is mimicked by 
Skyrmions for $B=3$ and 4.
These Skyrmions were studied in \cite{Leese:1994mc} using the
Atiyah-Manton ansatz \cite{Atiyah:1989dq}. 
It was observed that there is no negative
baryon density in the approximate 4-Skyrmion, but there is in the
approximate 3-Skyrmion case. In the approximate 3-Skyrmion there is a
region of negative baryon density surrounding the origin. This extends
out along four thin tubes which twice pinch to points and then widen
at very large distance until they merge and form another region of
negative baryon density at spatial infinity.

In this paper, we generalize the rational map ansatz so
that there can be negative baryon density. We calculate ansatz fields
that approximate the true minima more closely than the original
rational map ansatz. The ansatz for the 3-Skyrmion has
tubes of negative baryon density extending out from the origin; the
ansatz for the 4-Skyrmion does not.  Furthermore, there
is an octahedrally symmetric $B=5^*$ saddle-point configuration. The
ansatz for this saddle point also has negative baryon
density.

Thus, our investigation adds to the evidence that there may be regions
of negative baryon density in certain Skyrmions. This occurs in
those examples where the corresponding monopole has negative
multiplicity Higgs zeros. Of course, our conclusions are based on
an ansatz and the true solution does not necessarily possess the
same singularity structure. Unfortunately, it is difficult to observe
negative baryon density directly in the numerical solutions.

\subsection{Singularities of Differentiable Maps}
\label{singularities}

The theory of singularities deals with smooth maps between
manifolds. One of its main aims is to classify the points where the
Jacobian of a map does not have maximal rank. 
These are the singularities. Some
singularities are unstable, in the sense that a small perturbation
of the map can alter the nature of the singularity. For maps between low
dimensional manifolds, there is only a small number of stable
singularities. In this section, we describe the three stable
singularities of smooth maps between three-dimensional manifolds. We
will follow \cite{Arnold:1985}.

Let $f: M \to N$ be a map from a
three-dimensional manifold $M$ to a three-dimensional manifold
$N$. Locally, there are coordinates $\{y_1,y_2,y_3\}$ on $N$ and
$\{x_1,x_2,x_3\}$ on $M$ so that
\begin{eqnarray}
y_1&=&f_1(x_1,x_2,x_3),\nonumber\\
y_2&=&f_2(x_1,x_2,x_3),\nonumber\\
y_3&=&f_3(x_1,x_2,x_3).
\end{eqnarray}
The matrix $J=(\partial f_i/\partial x_j)$ is the Jacobian matrix of
the map. The singularities are the points where det$\,J=0$.  There are
only three stable singularities: folds, cusps and
swallowtails. These are described by giving their normal forms. The
normal form is a standard choice of coordinates for the neighbourhood
of the singularity. Any stable singularity can be expressed locally in
terms of the corresponding normal form by a smooth change of variables.

The simplest singularity is the fold, which can be visualized as the line
along which a piece of paper has been folded. A fold has the normal form:
\begin{eqnarray}
\label{efold}
  y_1&=&x_1^2,\nonumber\\ y_2&=&x_2,\nonumber\\ y_3&=&x_3.
\end{eqnarray}
It is worth considering the number of preimages of the map.  For
points of $N$ with $y_1 > 0$ there are two preimages, whereas for
points with $ y_1 < 0$ there are no preimages.  The fold is located at
$y_1 = 0$, which has one preimage. Restricted to the set of points $y_1
= 0$, $f$ maps the $x_2x_3$-plane onto the $y_2y_3$-plane.

The Jacobian matrix of this map is:
\begin{eqnarray}
J = 
\left(
\begin{array}{ccc}
2 x_1         & 0   & 0\\
0             & 1   & 0 \\
0             & 0   & 1 
\end{array}
\right).
\end{eqnarray}
It is singular at $x_1=0$, which, of course, is the location of the
fold. The rank of the Jacobian matrix at the fold is two. In fact,
this is true of all three stable singularities. Any singularity with a
rank one or rank zero Jacobian matrix is unstable.

Two folds can end on a cusp. This has the normal form:
\begin{eqnarray}
\label{ecusp}
  y_1&=&x_1^3 + x_1 x_2,\nonumber\\ y_2&=&x_2, \nonumber\\y_3&=&x_3.
\end{eqnarray}
In order to get a better understanding of this singularity, we
calculate the Jacobian matrix:
\begin{eqnarray}
J = 
\left(
\begin{array}{ccc}
3 x_1^2 + x_2 & x_1 & 0 \\
0             & 1   & 0 \\
0             & 0   & 1 
\end{array}
\right).
\end{eqnarray}
$J$ does not have maximal rank for $x_1^2 = - x_2/3$. This
is a pair of folds. The cusp occurs at the line $(0,0,x_3)$ where the
two folding surfaces $(\pm \sqrt{-x_2/3},x_2,x_3)$
meet. 

The most complicated stable singularity is called the swallowtail
and its normal form is
\begin{eqnarray}
\label{eswallow}
  y_1 &=& x_1^4 + x_1^2 x_2 + x_1 x_3,\nonumber\\ 
  y_2 &=& x_2,\nonumber\\ 
  y_3 &=& x_3. 
\end{eqnarray}
In this case, the Jacobian matrix is given by
\begin{eqnarray}
J = 
\left(
\begin{array}{ccc}
4 x_1^3 + 2 x_1 x_2 + x_3 & x_1^2 & x_1 \\
0             & 1   & 0 \\
0             & 0   & 1 
\end{array}
\right).
\end{eqnarray}
The points of the folds satisfy the equation $x_3 = -4 x_1^3 - 2 x_1
x_2$. The folds meet to form four cusp lines which meet at the
origin. The origin is the swallowtail.

This classification is known as Whitney's Theorem. This theorem states
that a map of a three-dimensional manifold to a three-dimensional
manifold is stable at a point if, and only if, the map can be described
in local coordinates in one of the four forms: a regular point with
$y_1=x_1$, $y_2=x_2$ and $y_3=x_3$ or one of the three singular forms
given above. Furthermore, maps with stable singularities are dense
in the space of all smooth maps: any map can be approximated
arbitrarily closely by a map with stable singularities.

\section{Folding and Rational Maps}
\label{folding}

We begin this section by showing that the simplest singularity of the
rational map ansatz is unstable. This will lead us to introduce the
non-holomorphic rational map ansatz in the following
section. Furthermore, we show that for $B>1$ there is an unstable
singularity at the origin.

In the holomorphic rational map ansatz there is a map from $\R^3$ to $S^3$ 
which maps $(r,z,{\bar z})$ to $(f(r), R(z), {\bar R} ({\bar z}))$.
Away from the origin, we can define local coordinates
$\{\mbox{Re}(z),\mbox{Im}(z),x_3\}$ and $\{y_1,y_2,y_3\}$ such that  
\begin{eqnarray}
y_1&=&\mbox{Re}(R),\nonumber\\ 
y_2&=&\mbox{Im}(R),\nonumber\\
y_3&=&x_3.
\end{eqnarray}
The simplest rational map with a singularity is 
\begin{equation}
\label{Rcomplex}
R(z) = z^2
\end{equation}
which gives
\begin{eqnarray}
y_1&=&x_1^2 - x_2^2,\nonumber\\ 
y_2&=&2 x_1 x_2,\nonumber\\
y_3&=&x_3.
\end{eqnarray}
The Jacobian matrix has rank one for the line $(0,0,x_3)$. This is not
one of the stable singularities. 
 
Let us consider small perturbations around (\ref{Rcomplex}). Adding
terms proportional to $z$ only shifts the singularity. Therefore, we
consider the following map:
\begin{equation}
\label{Rmod}
R(z,{\bar z}) =z^2 + 2 \epsilon {\bar z}.
\end{equation}
Since multiplying $\epsilon$ by a phase $e^{i \phi}$ only rotates the
singularities by $\phi$, we can take $\epsilon$ to be real. Using real
coordinates, the Jacobian matrix can be written as:
\begin{eqnarray}
J = 
\left(
\begin{array}{ccc}
2 x_1 + 2 \epsilon  & - 2 x_2 & 0 \\
2 x_2             & 2 x_1 - 2 \epsilon & 0 \\
0             & 0   & 1 
\end{array}
\right).
\end{eqnarray}
The Jacobian matrix is singular for $x_1^2 + x_2^2 = \epsilon^2$. 
Therefore, the singular points lie on a cylinder with radius $\epsilon$.
They can be parametrized by $x_1 = \epsilon \cos \alpha$ and $x_2 =
\epsilon \sin \alpha$ for $\alpha \in [0, 2 \pi]$, $x_3$
is arbitrary.
Restricting  the map to the singular surface, labeled by
$\alpha$ and $x_3$, we can calculate the cusp lines. 
The surface is singular where 
\begin{equation}
\frac{d y_1}{d\alpha}=\frac{d y_2}{d \alpha}=\frac{d y_3}{d \alpha}=0,
\end{equation}
therefore, the cusps form lines where $\alpha$ is zero, $2\pi/3$ or
$4\pi/3$, and $x_3$ is arbitrary.

\begin{figure}[!htb]
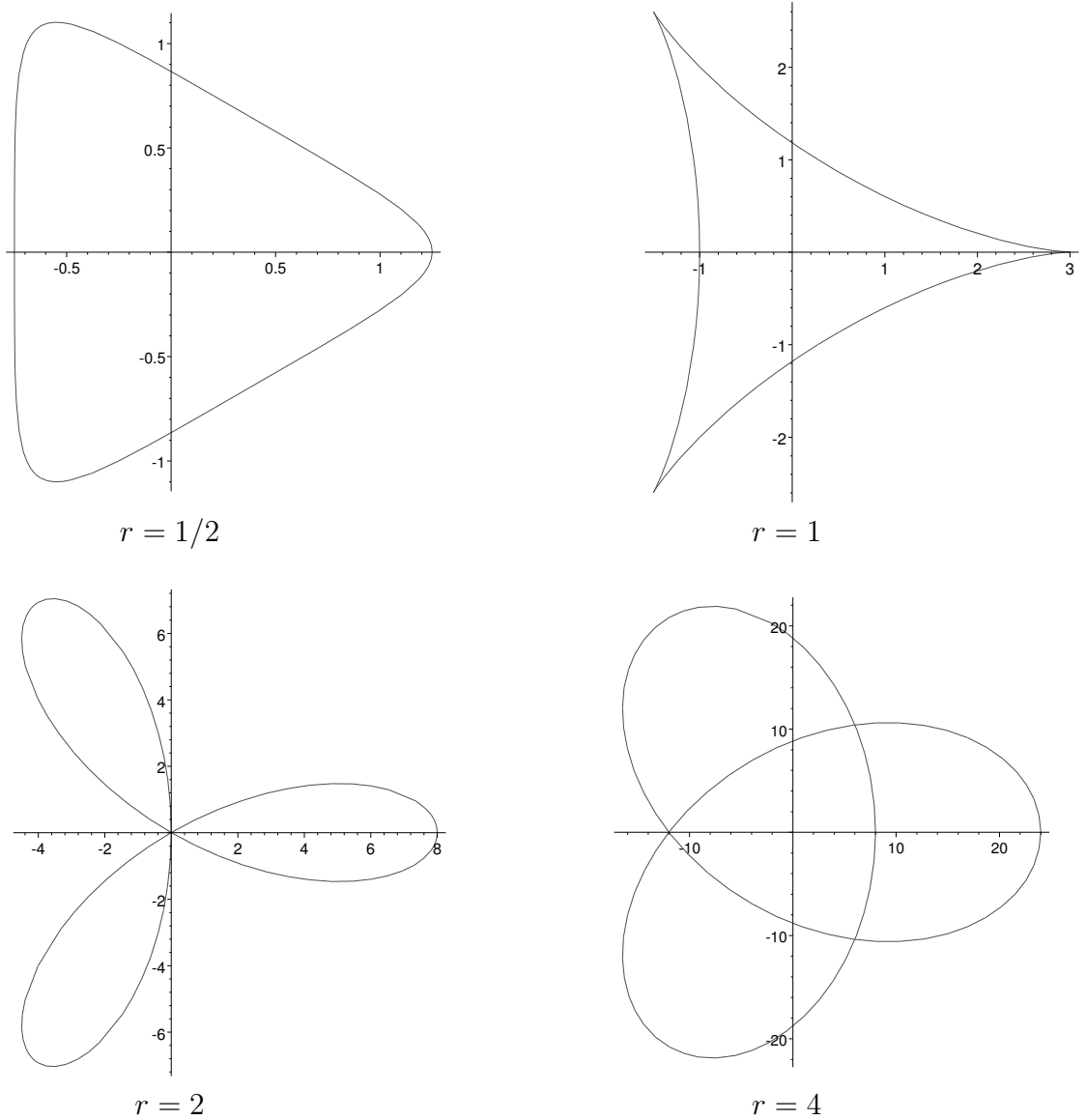

\begin{center}
\begin{tabular}{cc}
\hskip 7pt \quad\qquad\epsfxsize=6cm\epsffile{Krusch1a.epsi}
&\quad\qquad\qquad\begin{tabular}{c}{}\\[-192pt]
\epsfxsize=6cm\epsffile{Krusch1b.epsi}\end{tabular}\\
$r=1/2$&$r=1$\\ &\\
\qquad\qquad\epsfxsize=6cm\epsffile{Krusch1c.epsi}
&\quad\qquad\begin{tabular}{c}{}\\[-196pt]
\epsfxsize=6cm\epsffile{Krusch1d.epsi}\end{tabular}\\
$r=2$&$r=4$\\
\end{tabular}
\caption{The image of concentric circles $z = \rho {\rm e}^{i \phi}$ of
various radii $\rho$ under $R(z,{\bar z}) = z^2+2 {\bar z}$. Note that
the scale is different for each graph. 
\label{ffig0}}
\end{center}
\end{figure}

In Fig. \ref{ffig0}, we show the image of a set of concentric circles
of radius $\rho$ in the $x_1x_2$-plane. By rescaling space and target
space coordinates, the value of $\epsilon$ can be changed. For
convenience, we set $\epsilon = 1$ in the figure.  For small radius,
$\rho$, the ${\bar z}$ term is dominant and the image of the circle is a
deformed circle. As the radius increases, the circle becomes more
and more deformed. The $\rho=1$ circle maps to the singular
curve. This curve has three spikes. The points of these spikes are the
cusps and running between them are three folds.
Above this value of $\rho$, the map folds back on itself. The points
inside the fold have four preimages: 
a $\rho<1$ preimage with negative Jacobian and three $\rho>1$
preimages with positive Jacobians. Eventually, the image circle passes
completely through the folding region: for $\rho>3$ the image is a
trefoil shape. Every point outside the fold has just two preimages,
each with positive Jacobian. 

In general, the map $R(z) = z^n$ for $n \ge 2$ has an unstable singularity
on the $x_3$-axis. This unstable singularity can be removed by adding an
anti-holomorphic perturbation. The perturbed map $R(z) = z^n + n
\epsilon {\bar z}$ has a folding surface at $z = \epsilon \exp(i \alpha)$.   
There are $n+1$ cusps which are located at 
$\alpha = 2 \pi k/(n+1)$ where $k =0, \ldots, n$. 
This map possesses a natural $C_{n+1}$-symmetry, 
which maps the cusps into each other. 

It is worth noting that $z^2$ is a stable singularity within the set
of holomorphic functions, whereas $z^n$ for $n>2$ is unstable even as
a holomorphic map. Therefore, we do not expect the latter
singularities to occur for a generic holomorphic map.

Let us also consider the singularities of
the rational map ansatz at the origin. Locally, the shape function can
be written as  
\begin{equation}
f(r) = \pi - A r^\nu + O(r^{\nu+1}),
\end{equation}
where $A$ is an arbitrary positive constant and
\begin{equation}
\nu = -\frac{1}{2} + \frac{1}{2} \sqrt{8 B + 1}.
\end{equation}
For $B=1$ this exponent is equal to one. This means that the Jacobian
has full rank at the origin. However, for $B > 1$ the 
exponent $\nu$ is greater than one. In this case the Jacobian of the
rational map ansatz has rank zero. Moreover, the derivative $R_z(z)$
vanishes for certain values of $z$. Therefore, 
lines of degenerate singularities with rank one Jacobian matrices meet
at the origin. Here they form an even more degenerate singularity
with a rank zero Jacobian matrix. In Sect. \ref{tetrahedron}, we
argue that for $B=3$ there is a more likely singularity configuration
close to the origin.

\subsection{Non-Holomorphic Rational Maps}
\label{antiholomorphic}

The rational map ansatz restricts the ansatz fields in three
different ways. It requires that concentric two-spheres around the
origin in space are mapped to two-spheres in SU$_2$. 
It also requires that this map is the same
for all concentric two-spheres, and it requires 
that the map between two-spheres is a holomorphic map. Our
intention here is to relax the third of these conditions.
We will consider a larger class of maps. The maps we will consider will be
rational maps in the sense that they will have the form of a ratio between
polynomials. However, the polynomials will depend on $\bar{z}$ as well as
$z$, so they will not be holomorphic maps. 

To begin with, we consider the ansatz
\begin{equation}
U(r,z,\bar{z})=\exp{(i f(r)~\widehat{{\bf n}}_R\cdot \sigmabf)},
\end{equation}
where, as before,
\begin{equation}
\widehat{{\bf n}}_R=\frac{1}{1+|R|^2}(2\mbox{Re}(R), 2\mbox{Im}(R), 1-|R|^2)
\end{equation}
but, without assuming $R$ is holomorphic, 
\begin{equation}
R=R(z,\bar{z}).
\end{equation}
In order to derive the energy, $E$, we calculate the eigenvalues
$\lambda_1^2$, $\lambda_2^2$ and $\lambda_3^2$ of the strain tensor
(\ref{straintensor}). The strain tensor can be written as 
\begin{equation}
\left(D_{i j} \right) = 
\left(
\begin{array}{ccc}
{f^\prime}^2 & 0 & 0 \\
0 &
A \left(R_z+R_{\bar z} \right)\left({\bar R}_z+{\bar R}_{\bar z}
\right)  
&
i A \left(R_z {\bar R}_{z} - R_{\bar z} {\bar R}_{\bar z}
\right) \\
0 &
i A \left(R_z {\bar R}_{z} - R_{\bar z} {\bar R}_{\bar z}
\right)
&
- A \left(R_z-R_{\bar z} \right)\left({\bar R}_z-{\bar R}_{\bar z}
\right)
\end{array}
\right),
\end{equation}
where 
\begin{equation}
A = \left(\frac{1+|z|^2}{1+|R|^2} \frac{\sin f}{r} \right)^2,
\end{equation}
and has eigenvalues
\begin{eqnarray}
\nonumber
\lambda^2_1 &=&  {f^\prime}^2, \\
\nonumber
\lambda^2_2 &=& 
\left(
\frac{|R_{z}| + |R_{{\bar z}}|}{r} 
\frac{1 + |z|^2}{1 + |R|^2}
\sin f
\right)^2, \\
\lambda^2_3  &=& 
\left(
\frac{|R_{z}| - |R_{{\bar z}}|}{r} 
\frac{1 + |z|^2}{1 + |R|^2}
\sin f
\right)^2.
\end{eqnarray}
Notice that $\lambda^2_2$ and $\lambda^2_3$ are only equal if $R_{\bar z}
= 0$, or $R_{z} = 0$. In the first case, the energy of the holomorphic
rational map ansatz (\ref{Erational}) is reproduced. The second case
corresponds to a purely anti-holomorphic ansatz and is just the
holomorphic ansatz composed with a reflection.

Using equation (\ref{Egeo}) the energy $E$ can be rewritten as 
\begin{equation}
E = 4 \pi \int 
\left(
 r^2 f^{\prime 2} + 2 {\cal J} (f^{\prime 2} + 1) \sin^2 f 
 + {\cal I} \frac{\sin^4 f}{r^2}  
\right)
{\rm d}r,
\label{Eanti}
\end{equation}
where ${\cal I}$ and ${\cal J}$ are given by:
\begin{equation}
\label{calJ}
{\cal J} =\frac{1}{4 \pi} \int \left(
       \frac{(|R_z|^2 + |R_{{\bar z}}|^2)(1 + |z|^2)^2}{(1+|R|^2)^2} 
\right) 
\frac{2i~{\rm d}z {\rm d}{\bar z}}{(1 + |z|^2)^2}
\end{equation}
and
\begin{equation}
\label{fcalI}
{\cal I} = \frac{1}{4 \pi} \int \left(
      \frac{(|R_z|^2 - |R_{{\bar z}}|^2)(1+ |z|^2)^2}{(1+|R|^2)^2}    
\right)^2 \frac{2i~{\rm d}z {\rm d}{\bar z}}{(1+ |z|^2)^2}.
\end{equation}
As in the holomorphic ansatz, 
${\cal I}$ is essentially the integral of the square of the
Jacobian. However, ${\cal J}$ is not the integral of the Jacobian of $R$.

There is a close relationship between this energy functional and the
baby Skyrme model on a two-sphere.  The baby Skyrme model
\cite{Piette:1992dq, Piette:1995mh}
is a sigma-model in (2+1) dimensions with a fourth order
Skyrme-like interaction. The baby-Skyrme model on the two-sphere is of
independent interest, primarily because there is a phase transition as
the radius of the sphere is changed
\cite{Scoccola:1998ky,deInnocentis:2001ur}.

On a unit two-sphere, the baby Skyrme fields map $S^2$ to $S^2$
and the energy functional is of the form
$g_1{\cal J}+g_2{\cal I}$ where $g_1$ and $g_2$ are coupling
constants. ${\cal J}$ is the sigma-model energy and ${\cal I}$ is
the Skyrme energy. For a given shape function, $f$, the energy $E$ in
(\ref{Eanti}) is of this form with $g_1$ and $g_2$ calculated by
integrating the shape function over $r$.

Obviously, it would be possible to regard the holomorphic maps between
two-spheres as ans\"atze for baby-Skyrme fields. For a holomorphic map,
${\cal J}$ is equal to the topological degree. Thus, the holomorphic map
which minimizes ${\cal I}$ gives the best approximation to the energy
minimizing baby-Skyrme field. In other words, to find the baby-Skyrme 
energy minimizing holomorphic map, the values of $g_1$ and $g_2$ need
not be known. This is one of the reasons why the original
rational map ansatz is so convenient to use: the rational map is found
first and the shape function is then determined numerically.

For more general maps we need to employ an iterative
algorithm. Firstly, using the ${\cal I}$ minimizing holomorphic
rational map, a shape function can be 
calculated numerically. This gives provisional values for the coupling
constants $g_1$ and $g_2$. The next step is to minimize the
baby-Skyrme energy with these values of $g_1$ and $g_2$. This
determines a new map between the Riemann spheres.  The original shape
function is not optimized for this new map, so a new shape function
must be calculated. A new profile function gives new
coupling constants and so the whole procedure has to be iterated. In
practice, this procedure is simplified by the fact that we only
consider one-parameter families of non-holomorphic rational maps.

Another major advantage of using holomorphic maps is that there is
only a finite-dimensional family of holomorphic maps of given
topological degree. In contrast, the space of general maps between
two-spheres of a given degree is infinite dimensional. 
One way to avoid this problem would be to minimize the
baby-Skyrme model numerically. However, our interest here is in the
folding behavior of minimum energy Skyrmions and so we would prefer
to find an approximate analytic solution whose folding behavior we
can analyze. For this reason, we will restrict our attention to maps
of the form
\begin{equation}
R(z,\bar{z})=\frac{p(z,\bar{z})}{q(z,\bar{z})},
\end{equation}
where $p$ and $q$ are polynomials in both $z$ and $\bar{z}$. We will
call the polynomial degree of this map $(N_1,N_2)$, where $N_1$ is the maximal
holomorphic degree of the polynomials $p$ and $q$ and $N_2$ is their
maximal anti-holomorphic degree. By counting the number of preimages
of a given value of $R$ and taking into account the sign of the
Jacobian at each preimage, it follows that a general map of this form 
has topological degree $N_1-N_2$. Some maps will have a different
degree because $p$ and $q$ may have a common factor. Maps of this type
are called spurious.

According to formula (\ref{Bgeo}), the baryon number is
\begin{equation}
\label{Bantic}
B = \frac{2}{\pi} \int 
f^{\prime} \sin^2 f~ {\rm d} r~
\frac{1}{4 \pi}  \int \frac{(|R_z|^2 - 
          |R_{{\bar z}}|^2)(1+|z|^2)^2}{(1+|R|^2)^2} 
     \frac{2 i~ {\rm d}z {\rm d} {\bar z}}{(1 + |z|^2)^2}.
\end{equation}
$B$ is no longer an integral over squares, as it was for the
holomorphic rational map ansatz. It is possible that the baryon
density could be locally negative. We will find that this is what
happens for certain values of $B$.

\subsection{Symmetric Non-Holomorphic Rational Maps}

The Skyrmions that we are interested in are symmetrical: the 3-Skyrmion
is tetrahedrally symmetric and the $B=4$ is octahedrally symmetric
\cite{Braaten:1990rg}. Rather than minimizing the ansatz energy over
the whole space of maps, we restrict our attention to maps that have
the same symmetry as the numerically determined minimum energy
solution. 

An SO$_3$ rotation $g$ acts on the Riemann sphere $z$ as a M\"obius
transformation
\begin{equation}
z \mapsto g(z)=\frac{\alpha z + \beta}{-{\bar \beta} z + {\bar \alpha}},
\end{equation}
where $|\alpha|^2 + |\beta|^2 = 1$. There is also a M\"obius action on
the rational maps. A M\"obius transformation of the rational map is
equivalent to a global group transformation of the corresponding
Skyrme fields. A Skyrme field is symmetric under a rotation, $g$, if the
rotated fields are a global group transformation of the original fields. In
the same way, the rational map $R(z, {\bar z})$ is symmetric under $g$
if
\begin{equation}
\label{Rinv}
R \left(g(z),{\overline {g(z)}} \right)
= \frac{\alpha' R(z,\bar{z}) + \beta'}{-{\overline \beta'} R(z,\bar{z}) +
{\overline \alpha'}},
\end{equation}
where $\alpha'$ and $\beta'$ are not necessarily the same as $\alpha$
and $\beta$.

Symmetric non-holomorphic rational maps
can be calculated using elementary representation theory.  In this
subsection, we describe this construction. In the next section, we
will derive non-holomorphic rational maps for various $B$.

The Riemann sphere is isomorphic to $\CP^1$, the
one-dimensional complex projective space. $\CP^1$
can be labeled by a pair of complex numbers $[u,v]$, where the square
bracket indicates the relation
\begin{equation}
\label{Prelation}
\left[u, v \right] \cong \left[\lambda u ,\lambda v \right]
\end{equation}
with $\lambda \in \C^\times$.  These homogeneous coordinates, $u$
and $v$, are related to the inhomogeneous coordinate by $z = u/v$. The 
rotation group acts linearly on the homogenous coordinates and, so, it
is easier to use these coordinates to describe the representation
theory. However, we 
switch to inhomogeneous coordinates whenever they are more convenient.
Similarly, we can label the Riemann sphere by the complex conjugates
of $u$ and $v$, $[{\bar u}, {\bar v}]$, also subject to
relation (\ref{Prelation}).  A non-holomorphic rational map takes the
form 
\begin{equation}
\label{mapR}
 R(u,v,{\bar u}, {\bar v}) = \left[
p \left(u,v,{\bar u}, {\bar v} \right), 
q \left(u,v,{\bar u}, {\bar v} \right)\right].
\end{equation}
This rational map must be well-defined under the relation 
(\ref{Prelation}).Therefore,  $p$ and $q$ have to be homogeneous: they
are of the form   
\begin{eqnarray}
\label{phomogeneous}
p \left(u,v, {\bar u}, {\bar v} \right) 
&=& \sum_{i=0}^{N_1} \sum_{j=0}^{N_2} a_{i j} u^i v^{N_1-i} {\bar u}^j 
{\bar v}^{N_2-j}, \\
q \left(u,v, {\bar u}, {\bar v} \right) 
&=& \sum_{i=0}^{N_1} \sum_{j=0}^{N_2} b_{i j} u^i v^{N_1-i} {\bar u}^j 
{\bar v}^{N_2-j}.
\end{eqnarray}
It should be noted that the topological degree does not depend on the value of
$N_1+N_2$, it only depends on their difference.  
Choosing $N_1$ and $N_2$ corresponds to a truncation of
the possible maps between Riemann spheres. This truncation is similar
to truncating a Fourier expansion. In this paper we will consider
non-holomorphic maps with $N_2$ equal one or two.

Under an SO$_3$ rotation about the unit vector ${\widehat {\bf n}}$ by an
angle $\theta$,
the $[u,v]$ coordinates transform by an SU$_2$ transformation $\exp{(
i (\theta/2) {\widehat{\bf n}} \cdot \sigmabf)}$.
The SO$_3$ action on the Riemann sphere $[u,v]$ can now be written as: 
\begin{eqnarray}
\nonumber
u &\mapsto& u^\prime = (a_0 + i a_3) u + (a_2 + i a_1) v, \\
\label{Ptrafo}
v &\mapsto& v^\prime = (- a_2 + i a_1) u + (a_0 - i a_3) v,
\end{eqnarray}
where $a_i = n_i \sin (\theta/2)$ and $a_0 = \cos (\theta/2)$. 
The coordinates  $[{\bar u}, {\bar v}]$ transform as the complex
conjugate of (\ref{Ptrafo}). Let $G$ be the double group of a finite
subgroup of SO$_3$.
The rational map $[p,q]$ is $G$ invariant if an
SU$_2$ transformation of $[u,v]$ and $[{\bar u}, {\bar v}]$ is
equivalent to an SU$_2$ transformation of $[p,q]$.

A homogeneous polynomial of degree $N$ in $z$ transforms as ${\bf
N+1}$, the $(N+1)$-dimensional irreducible representation of
SU$_2$. It follows that the homogeneous polynomial $p$ of degree
$N_1$ in $z$ and degree $N_2$ in $\bar{z}$ transforms as $({\bf
N_1+1})\otimes ({\bf N_2+1})$. This representation can be decomposed
into irreducible representations of some finite group $G$. These
decompositions can be calculated using the characters. Tables of these
decompositions can be found, for example, in \cite{Koster:1963}.

By decomposing the $({\bf N_1+1})\otimes ({\bf N_2+1})$ as a
representation of $G$ we can determine whether or not there is a $G$
invariant degree $(N_1,N_2)$ rational map. In fact, the rational map
$[p,q]$ can be $G$-invariant in two different ways. One possibility is
that  
\begin{equation}
({\bf N_1+1})\otimes({\bf N_2+1})|_{G} = E \oplus\;\mbox{other
irreducible representations of $G$},
\end{equation}
and $\{p,q\}$ form a basis for the
two-dimensional irreducible representation $E$. This means $p$
and $q$ are transformed into linear combinations of each other under
M\"obius transformations of $z$. Moreover, by a choice of basis, these
combinations are unitary. This is always possible, because every
representation of a finite group is equivalent to a unitary
representation, \cite{Serre:1993}.
The second possibility is that
\begin{equation}
({\bf N_1+1})\otimes({\bf N_2+1})|_{G} = A_1 \oplus A_2
\oplus\; \mbox{other irreducible representations of $G$},
\end{equation}
and $p$ is a basis for $A_1$, and $q$ is a basis of $A_2$. Here,
$A_1$ and $A_2$ are two different one-dimensional
representations of $G$.  
In this case, there is a one-parameter family of $G$-symmetric
rational maps: namely $R=[p,a q]$. The parameter $a$ can be chosen to
be real, because a M\"obius transformation of $R$ can change the phase
of $a$. 

Of course, there is also a one-parameter family when
\begin{equation}
({\bf N_1+1})\otimes({\bf N_2+1})|_{G} = 2E \oplus\;\mbox{other
irreducible representations of $G$}
\end{equation}
because, in this case, there is a one-parameter family of choices of an
$E$ inside $2E$.
In order to construct a basis of this one-parameter family, we can
construct a projector $P_{\alpha \beta}$. Given a representation $\rho$
and a two-dimensional unitary representation $\rho^{(2)}_{\alpha
\beta}$, the projector is given by
\begin{equation}
P_{\alpha \beta} = \frac{2}{|G|} \sum_{g \in G} \rho^{(2)}_{\alpha
\beta} (g^{-1}) \rho(g).
\end{equation}
For details of this construction, see \cite{Houghton:1998kg, Krusch:2001}. 
It is not always necessary to construct
projectors. In the $B=2$ case discussed below, the invariant map is
calculated by direct calculation and in the $B=3$ case it is derived
from other, previously known, examples.

\section{Skyrmions from Non-Holomorphic Rational Maps}

In this section, we use non-holomorphic rational maps to approximate
the Skyrmions with baryon number two to four. In each of these cases,
there is a one-parameter family of maps with the correct
symmetry. Once the approximating map is found, we can discuss the
folding structure. We also consider the $B=5^*$ octahedral
saddle point which also has a one-parameter family of symmetric maps.
It is more tractable than the 5-Skyrmion, because the 5-Skyrmion is
not very symmetrical.  

\subsection{$B=2$: the Torus}

For $B=2$, the holomorphic rational map which minimizes ${\cal I}$ is: 
\begin{equation}
\label{ratB2}
R(z) = z^2.
\end{equation}
It has the same $D_{\infty}$ symmetry as the true solution. There is an
axial symmetry
\begin{equation}
\label{axialsym}
R(e^{i\chi}z)=e^{2 i \chi} R(z)
\end{equation}
and a symmetry under rotation by $\pi$ around an orthogonal axis
\begin{equation}
\label{invsym}
R \left(\frac{1}{z} \right) = \frac{1}{R \left(z \right)}.
\end{equation}

The group theory methods discussed in the last section are not really
needed here. The most general $D_\infty$-symmetric map can be
calculated by writing out the general $(3,1)$ rational map and
applying the symmetries (\ref{axialsym}) and (\ref{invsym}). It is
\begin{equation}
R(z,{\bar z}) = \frac{a z^2( z {\bar z} +1) + b z^2 (z {\bar z} -1)}
{a(z {\bar z} +1) + b (- z {\bar z} + 1)}.
\end{equation}
The true solution also has a reflection symmetry. In the holomorphic
case, that reflection symmetry is $R({\bar z}) =
{\overline{R(z)}}$. If we impose the same symmetry for the
non-holomorphic map, then the parameters $a$ and $b$ have to be real,
up to a common phase. 
Moreover, since the pair $(a,b)$ and $(\lambda a, \lambda b)$ gives
rise to the same rational map for all $\lambda \in \C^\times$, 
we can set $a = \cos \theta$ and $b = \sin \theta$.  The
polynomials have been chosen such that the value $\theta = 0$
corresponds to the holomorphic rational $(2,0)$ map. $\theta$ is in
the range $[-\pi/2,\pi/2]$, because under $\theta \mapsto \theta + \pi$
both $\sin \theta$ and $\cos \theta$ change sign.

\begin{figure}[!hbt]
\begin{center}
\includegraphics[height=150mm,angle=270]{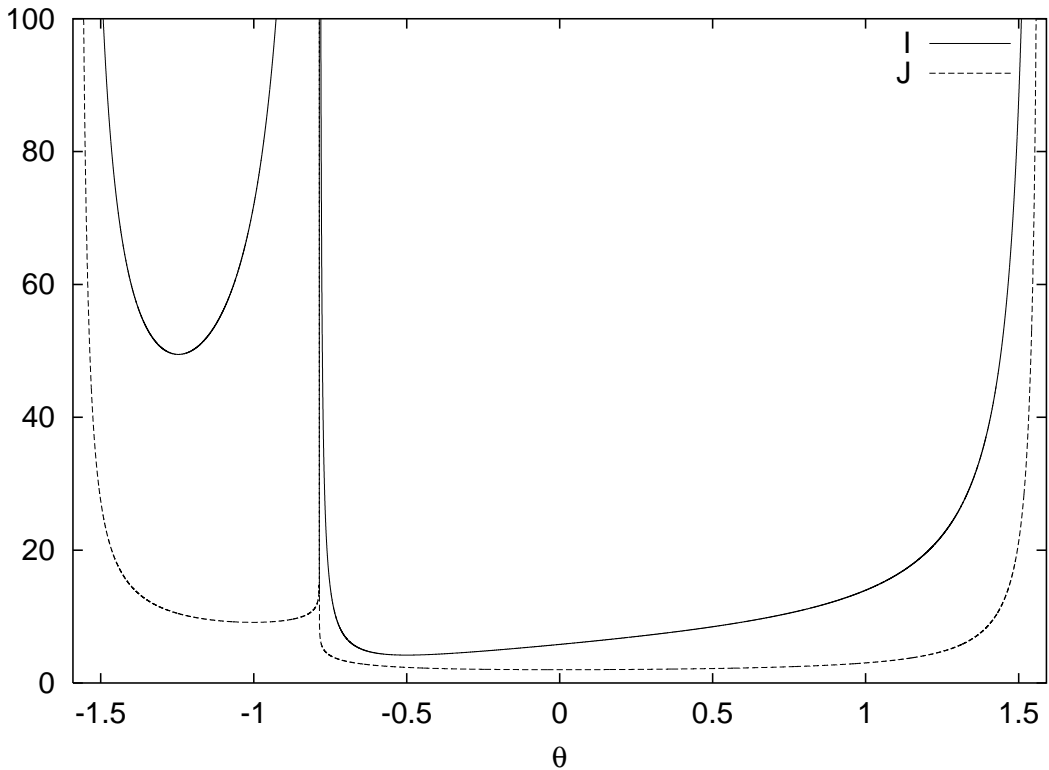}
\caption{${\cal I}$ and ${\cal J}$ as a function of $\theta$ for
$B=2$.\label{folding1}}
\end{center}
\end{figure}

In Fig. \ref{folding1}, we show the value of ${\cal J}$ and ${\cal
I}$ as a function of $\theta$. There are two poles, one at
$\theta=-\pi/4$ and another at $\theta=\pm \pi/2$. Both of these
poles correspond to points where the maps are spurious. At
$\theta=-\pi/4$, the cancellation of the common factor changes the
topological degree:
\begin{equation}
R(z,\bar{z})|_{\theta=-\pi/4}=\frac{z}{\bar{z}},
\end{equation}
whereas, at $\theta=\pm\pi/2$ it does not:
\begin{equation}
R(z,\bar{z})|_{\theta=\pm\pi/2}=-z^2.
\end{equation}
Thus, while the integrals are non-singular at $\theta=\pm\pi/2$, they 
diverge as this value of $\theta$ is approached. These
poles are considered in detail in \cite{Krusch:2001}. In
\cite{Krusch:2001}, it is also shown that there is some negative baryon
density only if $-\pi/2<\theta<-\pi/4$ or $\pi/4<\theta<\pi/2$.

To find the best approximation to the true minimum, we calculate the
value of $\theta$ which minimizes the energy.  The energy is a
combination of ${\cal I}$ and ${\cal J}$. The optimal value for ${\cal
J}$ is at the holomorphic rational map value $\theta =0$.  However,
${\cal I}$ is minimal for $\theta \approx -0.503$. Minimizing energy
(\ref{Eanti}) with respect to $\theta$ numerically, using the golden
rule algorithm \cite{NumericalRecipes:1992}, gives an optimal value of
$\theta \approx -0.202$. The energy per Skyrmion calculated with this
method is $E/B \approx 1.191$. In contrast, for the holomorphic
rational map $\theta = 0$, we obtain $E/B \approx 1.208$. 
The true value of the energy per Skyrmion is $E/B \approx 1.1791$. 
For the energy minimizing value of $\theta$, the baryon density is
positive everywhere except at the origin.  Thus, we find that there is
no negative baryon density for $B=2$, even though there is a
significant improvement in the energy. It seems that the axial
symmetry stabilizes the unstable singularity. 

\subsection{$B=3$: the Tetrahedron}
\label{tetrahedron} 

The 3-Skyrmion has tetrahedral symmetry $T$. The holomorphic rational
map ansatz is  
\begin{eqnarray}
\nonumber
p_T(z) &=&  - i \sqrt{3} z^2 + 1, \\
\label{poly30}
q_T(z) &=&  z^3 - i \sqrt{3} z,
\end{eqnarray}
and $\{p_T,q_T\}$ is a basis for the $E_2'$ in 
\begin{equation}
{\bf 4}|_T=E_2' \oplus E_3'.
\end{equation}
The decompositions can be easily derived from the
characters. Tables of these decompositions can also be found in, for example,
\cite{Koster:1963}. For convenience, a short table is given in Table
\ref{spinrepT}.

\begin{table}[tb]
\begin{eqnarray*}
{\bf 1}|_T&=&A_1\\
{\bf 2}|_T&=&E_1'\\
{\bf 3}|_T&=&F\\
{\bf 4}|_T&=&E_2' \oplus E_3'\\
{\bf 5}|_T&=&A_2 \oplus A_3 \oplus F\\
{\bf 6}|_T&=&E_1'\oplus E_2' \oplus E_3'\\
{\bf 7}|_T&=&A_1 \oplus 2F
\end{eqnarray*}
\caption{The decomposition of irreducible representations of SU$_2$
as representations of $T$.\label{spinrepT}} 
\end{table}

Here, we are interested in the non-holomorphic maps of degree
$(4,1)$. These correspond to the $10$-dimensional representation ${\bf
5}\otimes {\bf 2}$ and can be decomposed into
\begin{eqnarray}
 {\bf 5}\otimes {\bf 2} &=& {\bf 6}\oplus {\bf 4},\nonumber\\
 ({\bf 6}\oplus {\bf 4})|_T &=&
{E_1}' \oplus 2 {E_2}' \oplus 2 {E_3}'. 
\end{eqnarray} 
Both $\{1,z\}$ and $\{1, {\bar z}\}$ are a basis of the irreducible
representation ${E_1}'$ of $T$. When they are multiplied, they
decompose into $A_1 \oplus F$. For convenience, a multiplication table for
the tetrahedral group is given in Table \ref{Tmul}. 

\begin{table}[tb]
\begin{tabular}
{ccccccc|c}
$A_1$&$A_2$&$A_3$&$F$&$E_1'$&$E_2'$&$E_3'$&$\otimes$\\
\hline\\[-14pt]
$A_1$&$A_2$&$A_3$&$F$&$E_1'$&$E_2'$&$E_3'$&$A_1$\\
   &$A_3$&$A_1$&$F$&$E_2'$&$E_3'$&$E_1'$&$A_2$\\
   &   &$A_2$&$F$&$E_3'$&$E_1'$&$E_2'$&$A_3$\\
   &   &   &$A_1\oplus A_2\oplus A_3\oplus 
            2F$&$E_1'\oplus E_2'\oplus E_3'$ 
       &$E_1'\oplus E_2'\oplus E_3'$&$E_1'\oplus E_2'\oplus E_3'$&$F$\\
   &   &   & &$A_1\oplus F$&$A_2\oplus F$&$A_3\oplus F$&$E_1'$\\
   &   &   & &    &$A_3\oplus F$&$A_1\oplus F$&$E_2'$\\
   &   &   & &    &    &$A_2\oplus F$&$E_3'$
\end{tabular}
\caption{A multiplication table for the irreducible representations of $T$.
\label{Tmul}} 
\end{table}

A basis for this $A_1$ is:
\begin{equation}
\label{singlet}
k=z {\bar z} + 1.
\end{equation}
Furthermore, because  $A_1\otimes E_2'=E_2'$, $\{kp_T,kq_T\}$ is a basis for
an $E_2'$ inside ${\bf 5}\otimes {\bf 2}$. Explicitly, this basis is 
\begin{eqnarray}
p_1(z,{\bar z})&=& 
i \sqrt{3} z^3 {\bar z} + i \sqrt{3} z^2 - z {\bar z} - 1,
\nonumber\\ 
q_1(z, {\bar z}) &=& 
z^4 {\bar z} + z^3  - i \sqrt{3} z^2 {\bar z} - i \sqrt{3} z.
\end{eqnarray}
The rational map for this basis is spurious: $[p_1,q_1]=[kp_T,kq_T]=[p_T,q_T]$. The
common factor of $k$ cancels. However, when we have a second
independent basis, $\{p_2,q_2\}$, we will be able to form non-spurious
linear combinations involving $p_1$ and $q_1$.

To find an independent pair of basis vectors in $2E_2'$, 
we use the $A_2$ in ${\bf 5}|_T=A_2+A_3+F$. 
A basis for this $A_2$ is the Klein polynomial
\begin{equation}
k_f=z^4-2i\sqrt{3}z^2+1.
\end{equation}
This is often called the face polynomial, because its zeros are the
face points of a tetrahedron. Of course, the distinction between this
and the vertex polynomial is a matter of convention. Now, $A_2\otimes
E_1'=E_2'$ and so the required basis for $E_2'$ can be found by
multiplying $k_f$ by $[1,-\bar{z}]$. Thus, a second basis is given by 
\begin{eqnarray}
\nonumber
p_2(z,{\bar z}) &=& 
z^4 - 2 i \sqrt{3} z^2 + 1 \\
\label{specialpoly41}
q_2(z,{\bar z}) &=& 
-z^4 {\bar z} + 2 i \sqrt{3} z^2 {\bar z} - {\bar z}. 
\end{eqnarray}
As with the other basis pair, the rational map for this basis is
spurious:  
$[p_2,q_2]=[k_f,k_f\bar{z}]=[1,-\bar{z}]$. 

A general $T$-symmetric non-holomorphic rational map of degree $3$ is
given by 
\begin{equation}
\label{rat41}
R (z, {\bar z}) = 
\frac{\cos \theta p_1 + \sin \theta~ {\rm e}^{i \phi} p_2}
     {\cos \theta q_1 + \sin \theta~ {\rm e}^{i \phi} q_2}.
\end{equation}
The two angles $\theta$ and $\phi$ parameterize all maps. When
$\theta=0$ this map is spurious and reduces to the tetrahedrally
symmetric degree $(3,0)$ rational map studied in
\cite{Houghton:1998kg}. $\theta=\pm \pi/2$ is also spurious, here the map
reduces to a degree $(0,1)$ map. In this case, the cancellation
changes the topological degree of the map. It is expected that ${\cal
I}$ tends to infinity as this value of $\theta$ is approached.

For $\phi=0$, the map also has an additional reflection symmetry and
the symmetry group becomes $T_d$. Since the numerically determined
minimum seems to have this symmetry, it is expected that the best
result from the rational map ansatz will come from $\phi=0$. We have
confirmed this and will restrict our discussion to this case. 

\begin{figure}[!htb]
\begin{center}
\includegraphics[height=150mm,angle=270]{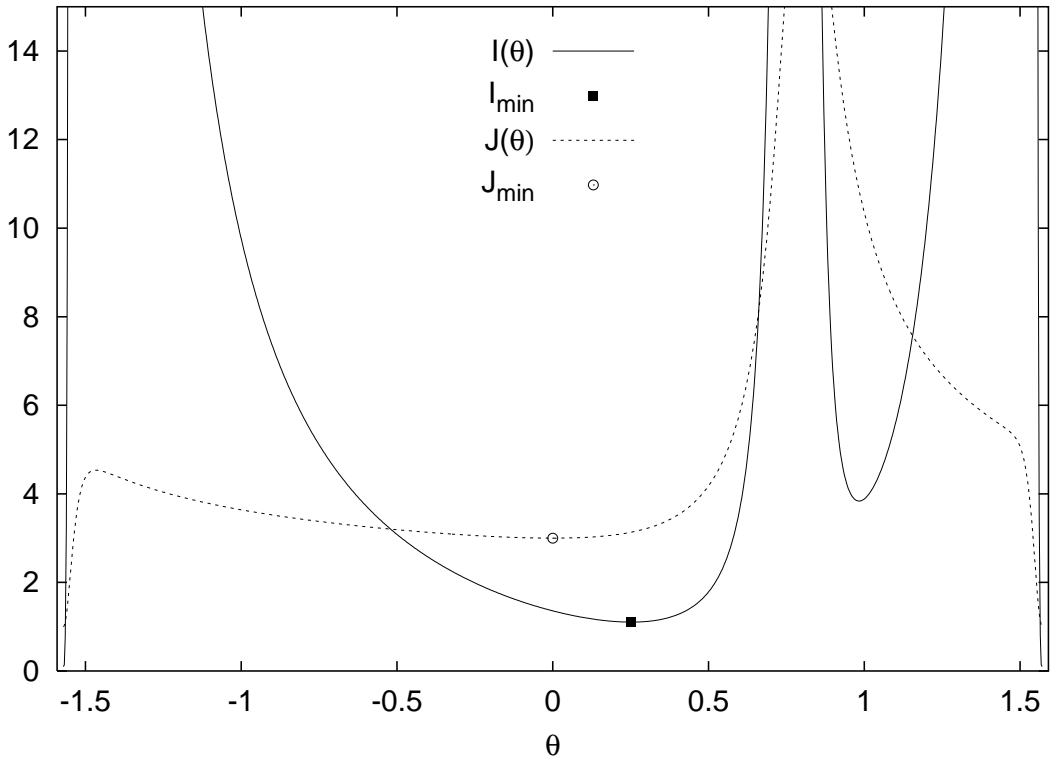}
\caption{${\cal I}$ and ${\cal J}$ as a function of $\theta$ for
$B=3$. Here ${\cal I}$ is rescaled by a factor of 10.\label{figB=3}}
\end{center}
\end{figure}

In Fig. \ref{figB=3}, we display ${\cal I}$ and ${\cal J}$ as a function
of the angle $\theta$. The minimum of ${\cal J}$ is still at the
rational map value $\theta = 0$ with ${\cal J} = 3$. However, the
minimum of ${\cal I}$ is at $\theta \approx 0.252$ with ${\cal I}
\approx 11.04$.  The minimum of the energy can now be calculated by
varying $\theta$, using a simple minimization scheme in which the
shape function is recalculate at each step. We obtain $E/B \approx
1.160$ for $\theta \approx 0.155$. 
The holomorphic rational map value is $E/B \approx 1.184$, 
whereas the exact solution has $E/B \approx1.1462$. Therefore, the
non-holomorphic rational map ansatz is a significant improvement. 

Furthermore, there are regions of negative baryon density. In the
holomorphic ansatz the singularities of the rational maps correspond
to points on the face centres of a tetrahedron.  
Let us consider the baryon density near these
points. The determinant of the Jacobian of the map $R$ is proportional
to the baryon density in (\ref{Bantic}):
\begin{equation}
\label{Bd}
B_R = \frac{(|R_z|^2 - |R_{\bar z}|^2)(1 + |z|^2)^2}{4 \pi (1 +|R|^2)^2},
\end{equation}
where we have integrated over $r$ using the boundary conditions on $f$.

It is convenient to reorient the map (\ref{rat41}) using the 
M\"obius transformation
\begin{equation}
z \mapsto \frac{2z + (\sqrt{3}-1)(1+i)}
{(1-\sqrt{3})(1-i) z + 2}.
\end{equation}
If the holomorphic map (\ref{poly30}) is rotated in this way, it is
singular at $z=0$, that is, $B_R$ for the holomorphic map
vanishes at $z=0$. 
If we rotate the non-holomorphic
map (\ref{rat41}) by the same M\"obius transformation and expand $B_R$
in terms of $z$ and ${\bar z}$ we obtain
\begin{equation}
B_R \approx - 0.17 + 10.8 z {\bar z},
\end{equation}
where $\theta$ has been set to the energy minimizing value
$0.155$. For $z=0$ the baryon density is negative and to lowest
order in $z$ and ${\bar z}$ the folds lie on a circle around the
origin in the $z$-plane. Thus, the non-holomorphic rational map
ansatz predicts tubes of negative baryon density.
In fact, the total negative baryon density, $B_-$, can be calculated
numerically. It is $B_- \approx 0.000089$.

Finally, we discuss the general singularity structure of the
3-Skyrmion.  All the singularities are of $z^2$ type and break up into
three cusps connected by folds. 
This is compatible with the tetrahedral symmetry. Therefore, a
3-Skyrmion consists of four tubes of folds, one through each of the
faces of the tetrahedron. Each of these tubes contains three cusp
lines. There are 12 cusp lines in total.

Unfortunately, it is difficult to directly examine the Skyrmion at the
origin and at infinity using these methods. It is possible to
speculate on what the singularity structure is, based on the
assumption that the singularities are all generic.

From the discussion in Sect. \ref{singularities} we know that four
cusp lines meet in a swallowtail. The cusp lines have to
respect the tetrahedral symmetry. If the singularities are generic,
they must meet in a swallowtail. Therefore, the 12 cusp lines cannot
meet at the origin but have to split up earlier.

Considering only the cusp structure, one possible configuration, would
be that the three cusp lines of each fold tube meet in a swallowtail,
resulting in one further cusp. The simplest tetrahedrally symmetric
configuration would be that this cusp meets similar cusps of the
remaining three fold tubes at the origin. However, we also know that
the instanton approximation to the 3-Skyrmion has negative baryon
density at the origin \cite{Leese:1994mc}, whereas this possible
configuration does not, because it has a singularity at the origin.

The following configuration is more likely. Each cusp of the fold
splits up into three cusps at a swallowtail at some distance from the
origin. Two of the cusp lines connect to the remaining two
swallowtails of the fold tube. The last cusp line connects to the
nearest swallowtail belonging to a different fold tube.  This
configuration is again tetrahedrally symmetric. Moreover, at the
origin the baryon density is non-zero.  
We will call the cusp lines which follow the
fold tubes {\it long cusp lines}, the cusp lines which connect swallowtails
of the same fold tube {\it short cusp lines} and the cusp lines which
connect swallowtails of different fold tubes {\it medium cusp lines}. In
this configuration there are 12 long cusp lines, six medium cusp lines,
and 12 short cusp lines. Note that the medium cusp lines lie on the
edges of a tetrahedron.

In order to decide the sign of the baryon
density at the origin, it is worth considering the folds of this
configuration. Folds separate positive from negative baryon
density. Moreover, precisely two folds end in one cusp. There are three folding
surfaces in each fold tube. Each of these folding surfaces ends in two
long cusp lines and one short cusp line. There are four additional folding
surfaces, which can be visualized as the sides of a tetrahedron. Each of
these folding surfaces ends in three medium cusp lines and three short
cusp lines. These are all the folding surfaces because precisely two
folding surfaces end in each of the cusps. Therefore, the baryon density
at the origin has the same sign as the baryon density inside the fold tubes.

Thus, the baryon density at the origin is negative, as it is in the
instanton ansatz, \cite{Leese:1994mc}. In the instanton configuration
the negative baryon density tubes pinch at two points. It is not
possible to decide using our methods whether this is a peculiarity of
the instanton ansatz or a feature of the 3-Skyrmion.

\subsection{$B=4$: the Cube}

The minimum energy $B=4$ configuration looks like a cube and  has
octahedral symmetry $O$. 
The corresponding invariant holomorphic map is
\begin{eqnarray}
\nonumber
p_O(z) &=&  2 \sqrt{3} z^2, \\
\label{poly40}
q_O(z) &=&  z^4  +  1.
\end{eqnarray}
These polynomials are a basis of the $E$ in
\begin{equation}
{\bf 5}|_O=E\oplus F_2.
\end{equation}
As before, these decompositions can be easily calculated, but, for
convenience, a table of them is given in Table \ref{Odecom}.

\begin{table}[tb]
\begin{eqnarray*}
{\bf 1}|_O&=&A_1\\
{\bf 2}|_O&=&E_1'\\
{\bf 3}|_O&=&F_1\\
{\bf 4}|_O&=&G'\\
{\bf 5}|_O&=&E\oplus F_2\\
{\bf 6}|_O&=&E_2'\oplus G'\\
{\bf 7}|_O&=&A_2\oplus F_1\oplus F_2\\
{\bf 8}|_O&=&E_1'\oplus E_2'\oplus G'\\
{\bf 9}|_O&=&A_1\oplus E\oplus F_1\oplus F_2
\end{eqnarray*}
\caption{The decomposition of the irreducible representations of SU$_2$
as representations of $O$.\label{Odecom}}
\end{table}

In the tetrahedral case, there was a unique invariant $(3,0)$ map and
a one-parameter family of $(4,1)$ maps. The same thing does not happen
here: if we consider $(5,1)$ maps we obtain
\begin{eqnarray}
{\bf 6}\otimes{\bf 2}&=&{\bf 7}\oplus{\bf 5}\nonumber,\\
({\bf 7}\oplus{\bf 5})|_O&=&A_2\oplus E\oplus F_1\oplus 2F_2,
\end{eqnarray}
and the $E$ is just the spurious map
$[kp_O,kq_O]=[p_O,q_O]$. $k=1+z\bar{z}$ is the invariant 
polynomial discussed earlier. In other words, the only invariant
$(5,1)$ map reduces to the holomorphic map.

This means that, in order to derive a one-parameter family of
invariant rational maps, we need to consider degree $(6,2)$:
\begin{eqnarray}
{\bf 7}\otimes{\bf 3}&=&{\bf 9}\oplus{\bf 7}\oplus{\bf 5},\nonumber\\
({\bf 9}\oplus{\bf 7}\oplus{\bf 5})|_O&=&A_1\oplus A_2\oplus 2E\oplus
2F_1\oplus 3F_2.
\end{eqnarray}
Thus, there is a one-parameter family corresponding to the $2E$.

Since ${\bf 3}\otimes{\bf 3}={\bf 5}\oplus {\bf 3}\oplus {\bf 1}$
there is an essentially unique degree $(2,2)$ SU$_2$ invariant. This
is $k^2$ and so $(k^2p_O,k^2q_O)$ spans an $E$ inside the
$2E$. Explicitly this is 
\begin{eqnarray}
p_1(z, {\bar z}) &=&
2 \sqrt{3} \left(
z^4 {\bar z}^2 + 2 z^3 {\bar z} + z^2
\right),
\nonumber\\
q_1(z, {\bar z}) &=&
z^6 {\bar z}^2 + 2 z^5 {\bar z} + z^4 + z^2 {\bar z}^2 + 2 z {\bar z}
+ 1.
\end{eqnarray}
By calculating the projection matrix we can derive another pair of
basis vectors in $2E$:
\begin{eqnarray}
\nonumber
p_2(z,{\bar z}) &=&
\sqrt{3} \left(
-z^6 + z^4 {\bar z}^2 - 8z^3 {\bar z} + z^2 -{\bar z}^2
\right),
\\
\label{specialpoly62}
q_2(z,{\bar z}) &=&
-z^6 {\bar z}^2 + 4 z^5 {\bar z} - 7 z^4 - 7 z^2 {\bar z}^2 + 4 z
{\bar z} -1,
\end{eqnarray}
and the general invariant rational map has the same form as in the
3-Skyrmion case, (\ref{rat41}). As before, $\phi = 0$ imposes the
reflection symmetry of the true solution, and $\theta=0$ corresponds
to the holomorphic map. 

\begin{figure}[!htb]
\begin{center}
\includegraphics[height=150mm,angle=270]{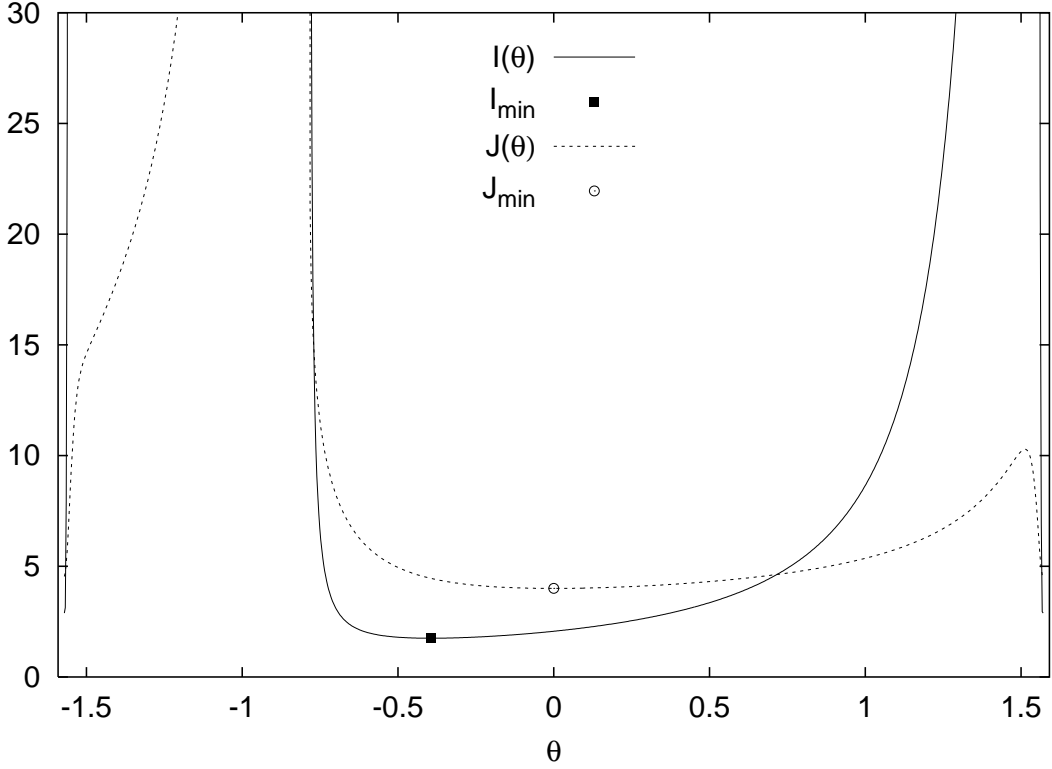}
\caption{${\cal I}$ and ${\cal J}$ as a function of $\theta$ for
$B=4$. Here ${\cal I}$ is rescaled by a factor of 10.\label{figB=4}}
\end{center}
\end{figure}

In Fig. \ref{figB=4} we show ${\cal I}$ and ${\cal J}$ as a function
of the angle $\theta$. At $\theta = 0$ the graph of ${\cal J}$ has 
its global minimum. There is a local minimum at $\theta = \pm \pi/2$.
The minimum value of ${\cal I}$ does not occur at the rational map
value but at $\theta \approx -0.40$. 
Therefore, there is the possibility of deriving a lower
energy. Indeed, minimizing the energy with respect to $\theta$ leads
to $E = 1.127$ for $\theta \approx -0.138$. This energy is only
$0.6\%$ above the true solution. By contrast the holomorphic rational
map ansatz energy is $1.5 \%$ above the true solution.

As before, we can examine the baryon density 
in the neighbourhood of a singularity of the holomorphic ansatz.
In this case, the map is already
oriented so that the holomorphic map has a singularity at
$z=0$. Setting $\theta = -0.138$ and expanding the density,
$B_R$, in powers of $z$ and ${\bar z}$ we obtain:
\begin{equation}
B_R \approx 2.54 z {\bar z}.
\end{equation}
Therefore, $B_R$ vanishes at $z=0$, and there is no negative baryon
density anywhere to lowest order. 

\subsection{$B=5^*$: the Octahedron}

The $B=5^*$ saddle point is octahedral in shape. The group theory involved
in this example is very like the group theory required for $B=3$. There is
a holomorphic map
\begin{eqnarray}
\nonumber
p_O(z) &=&  z^5 - 5 z, \\
\label{poly50}
q_O(z) &=&  - 5 z^4  +  1,
\end{eqnarray}
corresponding to the $E_2'$ in
\begin{equation}
{\bf 6}|_O=E_2'\oplus G'.
\end{equation}
The generalization to $(6,1)$ maps leads to a one-parameter family because
\begin{eqnarray}
{\bf 7}\otimes{\bf 2}&=&{\bf 8}\oplus{\bf 6},\nonumber\\
({\bf 8}\oplus{\bf 6})|_O&=&E_1'\oplus 2E_2'\oplus 2G'.
\end{eqnarray}
Multiplying the holomorphic maps by the SU$_2$ invariant, 
$k = z {\bar z} + 1$, gives
\begin{eqnarray}
p_1(z, {\bar z}) &=&
z^6 {\bar z} + z^5 - 5 z^2 {\bar z} - 5 z,
\nonumber\\
q_1(z, {\bar z}) &=&
- 5 z^5 {\bar z} - 5 z^4 + z {\bar z}
+ 1,
\end{eqnarray}
and calculating the projection matrices gives
\begin{eqnarray}
\nonumber
p_2(z,{\bar z}) &=&
3 z^6 {\bar z} - 23 z^5 -15 z^2 {\bar z} + 11 z,
\\
\label{specialpoly61}
q_2(z,{\bar z}) &=&
11 z^5 \bar{z} - 15 z^4 - 23 z {\bar z} + 3.
\end{eqnarray}
Thus, as before, a one-parameter family of invariant maps is given by
(\ref{rat41}) with $\phi = 0$. $\theta=0$ gives the holomorphic map. 

\begin{figure}[!htb]
\begin{center}
\includegraphics[height=150mm,angle=270]{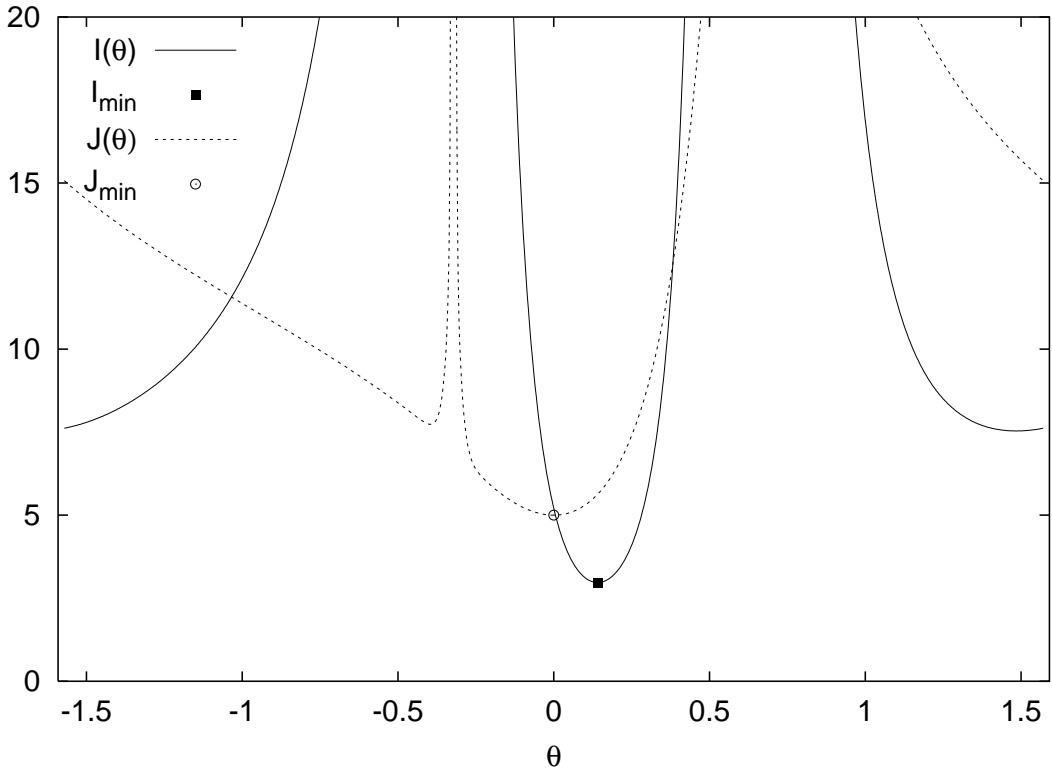}
\caption{${\cal I}$ and ${\cal J}$ as a function of $\theta$ for
$B=5^*$. Here ${\cal I}$ is rescaled by a factor of 10.\label{figB=5}}
\end{center}
\end{figure}

In Fig. \ref{figB=5} we display the graphs of ${\cal I}$ and ${\cal
J}$ as a function of $\theta$. There is a global minimum
of ${\cal J}$ at the rational map value ${\cal J} = 5$. The graph of
${\cal I}$ has a minimum at $\theta \approx 0.141$. Minimizing the
energy with respect to $\theta$ results in $E = 1.157$ for $\theta \approx
0.082$. This energy is only $1.7 \%$ above the true energy. In
comparison, the energy of the holomorphic rational map ansatz is 
$8.3 \%$ above the true energy. This significant improvement gives an 
indication that negative baryon density plays an important role in
this case. A local expansion shows that the non-holomorphic rational
map has negative Jacobian determinant
when $z$ takes the values of the singularities of the
holomorphic rational map. Finally, the integral of the negative baryon
density is $B_- \approx 0.00097$ which is a factor of ten larger than
for $B=3$.

\section{Conclusion}

This paper was motivated by the theory of singularities of
differentiable maps and, in particular, by Whitney's theorem. 
This theorem states that there
are only three types of stable singularities of maps between
three-manifolds.  
We showed that the holomorphic rational map ansatz for Skyrmions
\cite{Houghton:1998kg} does not have a stable singularity structure: it
does not even allow folding. We introduced a non-holomorphic rational map
ansatz that allows folding. For baryon numbers two, three and four,
the approximate solutions derived from this ansatz are closer in
energy to the true Skyrmions than any other ansatz solution.

The key idea of the non-holomorphic rational map ansatz is to
construct maps between Riemann spheres which are not holomorphic but
have the same symmetry as the true solutions. We described this
construction in detail and calculated the non-holomorphic rational
maps for baryon numbers two, three and four, and for the
octahedrally-symmetric $B=5^*$ saddle point. Non-holomorphic rational
maps can have folding, and therefore, negative baryon density. We
found that there is negative baryon density for $B=3$ and $B=5^*$ but
not for $B=2$ and $B=4$.

By decomposing group representations we showed that there is a
one-parameter family of tetrahedrally-symmetric $(4,1)$ maps. These
maps have topological degree three. The ansatz energy was minimized
within this family to find the best approximation to the
3-Skyrmion. The $(4,1)$ maps have anti-holomorphic degree one and so
they are minimal generalizations of the original holomorphic map. For
$B=5^*$ the situation is similar: there is a one-parameter family of
octahedrally symmetric $(6,1)$ maps. However, for $B=4$ the $(5,1)$
maps do not contain a one-parameter family of octahedral maps, only
the $(6,2)$ maps do.  The $(5,1)$ maps can be thought of as the first
order effect, and the $(6,2)$ maps as a second order effect. In fact,
for $B=4$, the holomorphic rational map is remarkably close to the
true solution. The energy error is $1.5 \%$. The holomorphic rational map
approximation to the 3-Skyrmion has an error of $3.3 \%$. 
$B=7$ seems to be similar
to $B=4$.  The 7-Skyrmion is dodecahedral and it is easy to check that
only the $(10,3)$ maps contain a one-parameter family of
icosahedrally-symmetric maps. 
Again, the holomorphic rational map is extremely close to the
true solution, with an error of only $1.1 \%$. Therefore, we do not
expect that negative baryon density occurs for $B=7$.

There is an icosahedrally-symmetric $B=11^*$ saddle point. The
holomorphic rational map ansatz approximates it quite poorly. It
predicts an energy which is far larger than eleven 1-Skyrmions,
whereas the true solution is a saddle-point solution with $E/B =
1.158$.  This can be viewed as an indication that negative baryon
density plays a major role. The representation theory also suggests
that there is negative baryon density because there is a one-parameter
family of icosahedrally symmetric $(12,1)$ maps.

It seems possible to decide heuristically whether or not a Skyrmion of
a certain symmetry possesses negative baryon density. In Sect.
\ref{singularities} we showed that a $z^n$ singularity can be decomposed
into folds which contain $n+1$ cusps and that there is a natural
$C_{n+1}$ symmetry which maps the cusps into each other. It seems
likely that negative baryon density occurs if this $C_{n+1}$ symmetry
is compatible with the symmetry of the faces.

Direct computation shows that in the holomorphic rational map ansatz
for $B=2$, $3$, $4$, $5^*$, $7$ and $11^*$ all the singularities are of
$z^2$ type. Therefore, negative baryon density occurs, if the faces
have a $C_3$ symmetry.  The faces of the tetrahedron, the octahedron
and the icosahedron are equilateral triangles. This is consistent with what we
found: there is negative baryon density for $B=3$ and $5^*$. It
suggests that there is also negative baryon density for $B=11^*$. On
the other hand, the faces of a torus are round, the faces of a cube are
squares and the faces of a dodecahedron are pentagons. Correspondingly,
we did not find any negative baryon density for $B=2$ and $4$ and do
not expect negative baryon density for $B=7$.

In the 3-Skyrmion case we discussed the singularity structure at the origin. 
Assuming tetrahedral symmetry and generic singularities
we conjectured a configuration with $16$ folding surfaces, $30$ cusp
lines and $12$ swallowtails. This is compatible with the instanton
calculations in \cite{Leese:1994mc}. 

It appears that there is a large number of swallowtails, cusps and folds
for both the 3-Skyrmion and $B=5^*$ saddle point, but non-generic
singularities for the 2-Skyrmion and 4-Skyrmion. Thus, while it may be
that in non-extremal Skyrme configurations there are certain folding
features associated with each interacting 1-Skyrmion, this is not
necessarily apparent in the extremal configurations.

It might be interesting to examine the number of anti-vacuum points, that
is, the number of points where $U = -1$.
A 1-Skyrmion is centered around a single anti-vacuum point and in a
configuration of well-separated Skyrmions, the individual Skyrmions
can be thought of as being positioned at the anti-vacuum
points. However, as the Skyrmions approach each other there may be
more anti-vacuum points, some with positive Jacobian and some with
negative. This is certainly what is implied by the folding seen in the
3-Skyrmion and $B=5^*$ saddle point, and by the corresponding monopole
configurations. 

The topological charge of a BPS monopole is equal to the number of zeros
of the Higgs field provided that the zeros are counted with their
multiplicity. Furthermore, well-separated 1-monopoles are centered
around a zero of the Higgs field. However, the total number of zeros can
exceed the topological charge. In \cite{Houghton:1996uj}, 
3-monopole fields were calculated
using the mixture of analytic and numerical methods first described in
\cite{Hitchin:1995,Houghton:1996bs}. 
It was found that the number of zeros of the Higgs field can be
as high as seven, with five zeros of positive winding number and two of
negative winding number. In \cite{Sutcliffe:1996qz}, the cubic
4-monopole, octahedral 5-monopole and dodecahedral 7-monopole
calculated in \cite{Hitchin:1995,Houghton:1996bs,Houghton:1996} were 
examined and it was found that zeros with negative winding number occur
for the octahedral 5-monopole but not for the cubic 4-monopole and the
dodecahedral 7-monopole. This pattern is mimicked by what we have found
for Skyrmions. This possibility was discussed in \cite{Sutcliffe:1996qz}.

It is also interesting that in monopole dynamics, as an individual
monopole approaches other monopoles,  its zero of the Higgs field
often splits into three zeros, two with positive winding and one with
negative winding \cite{Houghton:1996uj,Sutcliffe:1996qz}. Perhaps
something similar happens in the case of Skyrmions. 

\section*{Acknowledgments}

We would like to thank N S Manton for useful discussion and for drawing
our attention to the literature concerning the singularities of
differentiable maps.  One of us, S K thanks PPARC and the
Studienstiftung des deutschen Volkes for financial assistance.

\appendix
\section*{Appendix: Numerical Results}

For convenience the main numerical results have been gathered together
in Table \ref{nr}. As discussed above, the table shows that the minima
of ${\cal I}$ and ${\cal J}$ do not coincide and so the more general
non-holomorphic rational map ansatz is closer to the true energy than
the holomorphic ansatz. 
However, only for $B=3$ and $B=5^*$ are there regions of
negative baryon density. These regions are quite small.

\begin{table}[htb]
\begin{center}
\begin{tabular}{|c||c||c||c||c|}
\hline
&&&&\\
&True&Holomorphic&Minimizing ${\cal I}$&Minimizing
Energy\\
&&&&\\
\hline
\begin{tabular}{c}\\
$B$\\
\hphantom{$5^*$}\\
\end{tabular}&
\begin{tabular}{c}\hphantom{1.232}\\
$E/B$\\
\\
\end{tabular}&
\begin{tabular}{c|c}&\\ ${\cal I}$&$E/B$\\ \hphantom{52.10}
&\hphantom{1.232}\\
\end{tabular}&
\begin{tabular}{c|c}&\\
${\cal I}$&$B_-$\\
\hphantom{29.72}&\hphantom{0.00088}\\
\end{tabular}&
\begin{tabular}{c|c|c|c}&&&\\
${\cal J}$&${\cal I}$&$B_-$&$E/B$\\\hphantom{5.20}
&\hphantom{32.93}&\hphantom{0.00097}&\hphantom{1.157}\\ 
\end{tabular}\\
\hline
\begin{tabular}{c}1\\\\2\\\\3\\\\4\\\\
$5^*$\\
\end{tabular}&
\begin{tabular}{c}1.2322\\\\1.1791\\\\1.1462\\\\1.1201\\\\1.138\\
\end{tabular}&
\begin{tabular}{c|c}&\\1&1.232\\&\\5.81&1.208\\&\\13.58&1.184\\&\\20.65&1.137\\&\\52.10&1.232\\&
\end{tabular}&
\begin{tabular}{c|c}&\\1&0\\&\\
4.20&0\\&\\11.04&0.00088\\&\\17.50&0\\&\\29.72&0.00841\\&
\end{tabular}&
\begin{tabular}{c|c|c|c}&&&\\
1&1&0&1.232\\&&&\\
2.04 & 4.98 &0&1.191
\\&&&\\
3.04&11.48&0.00009&1.160\\&&&\\
4.04&18.95&0&1.127\\&&&\\
5.20&32.93&0.00097&1.157\\&&&
\end{tabular}\\
\hline
\end{tabular}
\end{center}
True: These are the energies obtained in
\cite{Battye:2000se}
by numerical minimization of the Skyrme energy.\\ 
Holomorphic: 
These are the values obtained in \cite {Houghton:1998kg} using
the holomorphic  ansatz.\\
Minimizing ${\cal I}$: ${\cal I}$ is minimized within the class of maps
considered.\\
Minimizing Energy: These are the best
results obtained with the non-holomorphic rational map ansatz.
\caption{The numerical results.\label{nr}}
\end{table}

\end{document}